\documentclass[letterpaper]{article}
\usepackage[preprint]{aaai2027}
\usepackage[hyphens]{url}
\usepackage{graphicx}
\usepackage{natbib}
\usepackage{caption}
\usepackage{booktabs}
\usepackage{multirow}
\usepackage{amsmath,amssymb}

\title{TeleAntiFraud 2.0: A Refreshable, Profile-Grounded, and\\
Audio-Based Benchmark for Telecom Fraud Detection}
\author{
    Huiyuan Liu\textsuperscript{1}\equalcontrib, Zhiming Ma\textsuperscript{2}\equalcontrib, Yanxing Liu\textsuperscript{3}, Shun Zhang\textsuperscript{1}\corresponding \\
    Qifan Wang\textsuperscript{4}, Di Liu\textsuperscript{1}, Yifan Wang\textsuperscript{1}, Yuyang Deng\textsuperscript{1} \\
    Haoyang Meng\textsuperscript{1}, Yijin Zhou\textsuperscript{5}, Yuxi Zhao\textsuperscript{1}, Chengxian Hu\textsuperscript{1} \\
    Peidong Wang\textsuperscript{6}, Peng Chen\textsuperscript{1}
}
\affiliations{
    \textsuperscript{1}People's Public Security University of China \quad
    \textsuperscript{2}JD Technology \\
    \textsuperscript{3}Chongqing Ant Consumer Finance Co., Ltd., Ant Group \quad
    \textsuperscript{4}Meta \\
    \textsuperscript{5}University of Science and Technology of China \quad
    \textsuperscript{6}Northeastern University \\[0.5ex]
    {\small
    \{202421250002, 2024111026, 202121710003\}@stu.ppsuc.edu.cn \\
    \{202121710017, 202421450003, 202421430031\}@stu.ppsuc.edu.cn \\
    \{shunzhang, liudi, chenpeng\}@ppsuc.edu.cn \\
    mazhiming312@outlook.com \quad liuyanxing21@mails.ucas.edu.cn \\
    wqfcr@meta.com \quad zyjm@mail.ustc.edu.cn \quad pdongwang@163.com}
}

\begin{document}

\maketitle

\begin{abstract}
Telecom fraud scripts evolve rapidly and are often designed to resemble routine service conversations, creating two key requirements for audio-based telecom-fraud evaluation. First, benchmarks must incorporate newly observed scam patterns without overwriting previously established test sets. Second, they must distinguish fraud from lawful, near-domain calls rather than relying on topic-separated negative examples. We present \textsc{TeleAntiFraud 2.0}, constructed with our Mixed-Tree Anti-Fraud Generation Pipeline and evaluated under a monthly frozen evaluation protocol. The pipeline transforms online fraud-case abstracts into profile-grounded scenarios, expands them through mixed-tree generation, realizes fraud and non-fraud dialogue paths under shared contexts, renders validated dialogues as role-matched speech, and freezes the resulting audio, labels, prompts, manifests, and provenance records for each monthly evaluation set. Each frozen set contains 900 Chinese calls, comprising 600 fraud and 300 near-domain non-fraud cases. Controlled text experiments show that three classifiers achieve perfect macro-averaged F1 (Macro-F1) when evaluated against unrelated or ordinary negatives, but drop to 0.65--0.68 with near-domain sibling negatives. Full-set audio and automatic-speech-recognition plus large-language-model (ASR+LLM) evaluations further reveal class-prior shortcuts, prediction collapse, and snapshot sensitivity. Together, these findings establish near-domain construction and collapse-aware reporting as core requirements for evaluating audio-based telecom-fraud models under realistic confusable conditions. The accompanying research artifact includes the construction code, evaluation scripts, manifests, and documentation. Our dataset and code are available at \url{https://anonymous.4open.science/r/TeleAntiFraud-2_0-EEB2/}.
\end{abstract}

\section{Introduction}

Audio-based telecom-fraud detection is a high-stakes speech and language understanding task with direct public-security implications. The 2024 Global State of Scams report estimates global scam losses of more than USD 1.03 trillion, based on 58,329 survey responses~\cite{gasa2024global}. Fraudulent calls combine identity impersonation, procedural framing, urgency, and coercion to persuade recipients to disclose sensitive information or take harmful actions. At the same time, scam scripts evolve rapidly and are often crafted to resemble legitimate customer-service, risk-notification, and verification calls. These characteristics make isolated keywords and early conversational cues insufficient for reliable detection. Instead, robust detection requires reasoning over the complete interaction, including the actions requested by the caller, whether independent verification remains possible, and how the conversation ultimately concludes.

\begin{figure}[t]
\centering
\includegraphics[width=1.00\linewidth]{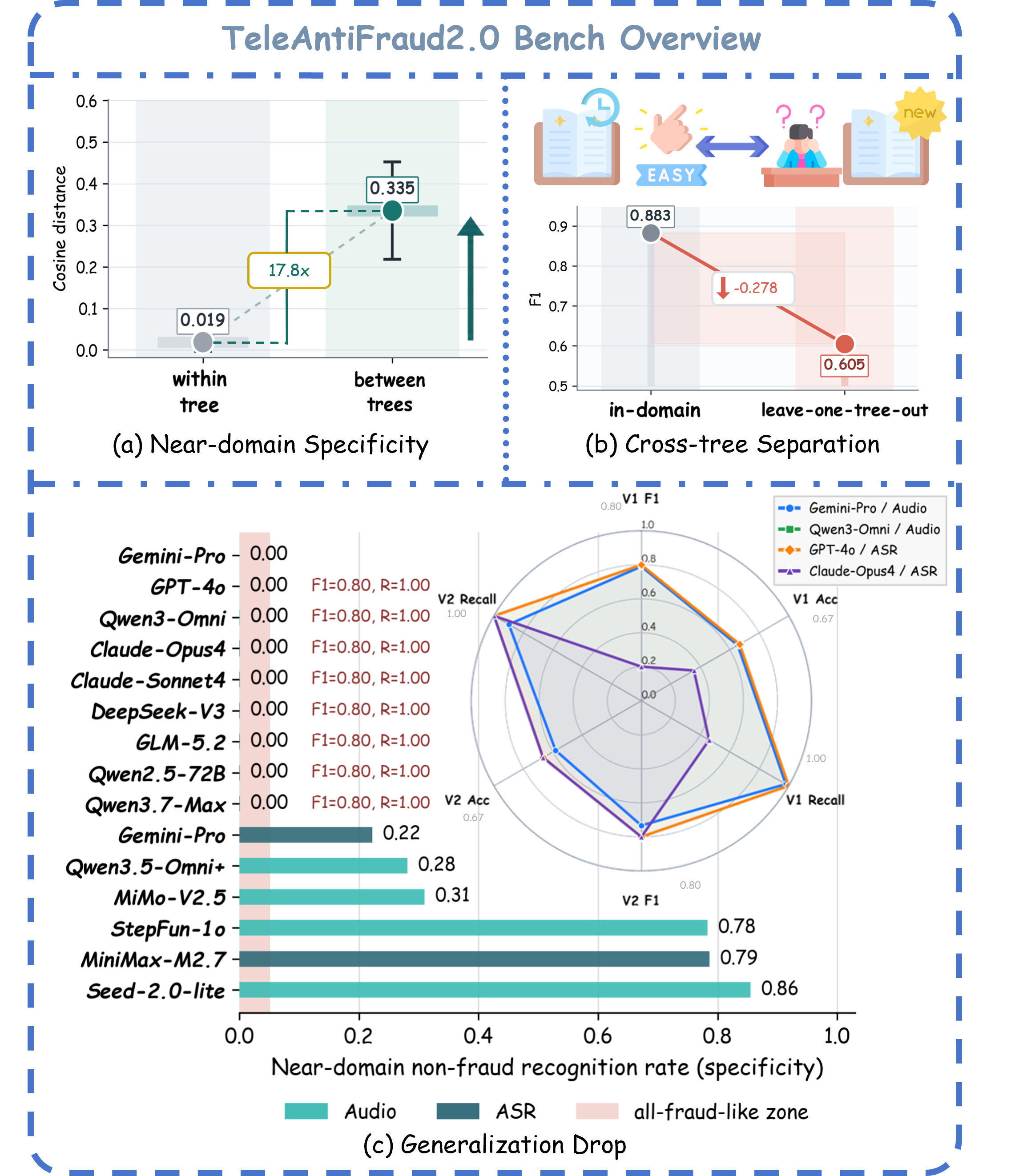}
\caption{Mixed-tree construction and snapshot-level diagnostics, including tree-distance structure, transfer difficulty, prediction collapse, and snapshot variation.}
\label{fig:intro-overview}
\vspace{-3mm}
\end{figure}
Fraud detection has a long history in statistical modeling, machine learning, and the analysis of social-engineering tactics~\cite{bolton2002statistical,kou2004survey,abdallah2016fraud,stajano2011scam,vishwanath2011phished}. Recent benchmarks have advanced the systematic evaluation of anti-fraud systems. The Fraud Dataset Benchmark (FDB) standardizes heterogeneous public fraud datasets through a unified interface~\cite{fdb}, while Fraud-R1 extends evaluation to multi-round fraud and phishing inducement scenarios~\cite{fraudr1}. For spoken telecom fraud, TeleAntiFraud-28k introduces an audio-text benchmark for slow-thinking analysis of fraudulent calls~\cite{teleantifraud}. These resources provide important foundations, but they do not jointly address two central challenges in spoken telecom-fraud evaluation: continuously refreshing benchmarks as fraud patterns evolve and distinguishing fraud from realistic, near-domain lawful calls.

Despite this progress, two gaps remain. First, fixed test sets cannot incorporate scam patterns observed after their release, even as impersonated institutions, requested actions, and persuasion strategies continue to change. Second, when non-fraud examples come from unrelated topics or data sources, models may rely on lexical or source-specific shortcuts rather than the actions that distinguish fraudulent from lawful calls. Continually replacing old test examples does not resolve these gaps, because a constantly changing evaluation set would make results difficult to reproduce and compare over time. Telecom-fraud evaluation therefore needs to absorb newly observed cases while preserving previously released test sets and their evaluation records.

We introduce \textsc{TeleAntiFraud 2.0}, a versioned audio benchmark organized as monthly frozen snapshots. Newly collected fraud case abstracts can be incorporated into later releases, while every published snapshot remains immutable. Each snapshot preserves the audio, labels and rationales, generation metadata, evaluation prompts, model responses, and provenance records needed to reproduce and audit its results. This design allows the benchmark to track newly observed scam patterns without overwriting prior evaluation sets.

To construct each snapshot, we develop the Mixed-Tree Anti-Fraud Generation Pipeline, which converts fraud case summaries into structured scenario profiles, expands the profiles into mixed dialogue trees, realizes dialogue paths through collaborative role-playing and functional agents, and renders validated dialogues as role-matched speech. Within each tree, fraud and lawful sibling paths share participants, scenario context, opening turns, and early risk language, and diverge only after label-bearing actions emerge. The resulting labels therefore depend on the completed interaction trajectory rather than on topic-level cues. Figure~\ref{fig:intro-overview} provides a compact overview of the benchmark construction and the diagnostic behaviors that motivate collapse-aware reporting.

Controlled text experiments show that unrelated and ordinary telecom negatives make the task nearly perfectly separable, whereas near-domain sibling negatives reduce Macro-F1 to 0.65--0.68. Full-set audio and ASR+LLM evaluations further show false-positive bias, class-prior shortcuts, prediction collapse, and substantial variation across monthly snapshots. These findings show that fraud-class F1 alone is insufficient for telecom-fraud evaluation and motivate joint reporting of class-balanced metrics, class-conditional recall, prediction distributions, and collapse behavior. The two current snapshots support snapshot-sensitivity analysis; longer-term temporal claims require additional releases.

Our contributions are threefold:
\begin{itemize}
    \item We introduce a \textbf{versioned audio benchmark for continuously evolving telecom fraud}. Monthly releases incorporate newly observed scam patterns while keeping every published snapshot immutable, together with the artifacts required for reproducible and auditable evaluation.
    \item We develop the \textbf{Mixed-Tree Anti-Fraud Generation Pipeline}. Fraud case summaries are transformed into mixed dialogue trees and role-matched speech, with fraud and lawful sibling paths sharing context and diverging only at actions that provide sufficient label evidence.
    \item We conduct \textbf{controlled text and full-set audio evaluations} showing that unrelated negatives substantially overestimate detection performance, whereas near-domain siblings expose false-positive bias, class-prior shortcuts, and prediction collapse across model families. The benchmark and its analysis protocol provide the community with a harder and more auditable testbed for tracking progress in audio-based telecom-fraud detection.
\end{itemize}

\section{Related Work}

\textbf{Anti-fraud datasets and benchmarks.} Fraud detection has long been studied through statistical modeling, data mining, and domain-specific machine learning~\cite{bolton2002statistical,kou2004survey,abdallah2016fraud}. Public benchmarks have extended this line by collecting fraud datasets under shared evaluation interfaces. FDB~\cite{fdb} aggregates public fraud datasets across domains and highlights class imbalance, heterogeneous features, temporal patterns, and adversarial behavior. Recent LLM-oriented benchmarks further move beyond static records: Fraud-R1~\cite{fraudr1} evaluates multi-round resistance to fraud and phishing inducements, including role-play settings. Table~\ref{tab:comparison} shows that anti-fraud evaluation must account for changing tactics and interactive persuasion, but existing benchmarks do not directly model spoken telecom calls.

\noindent\textbf{Dynamic and auditable evaluation.} Benchmark studies increasingly treat evaluation sets as maintained instruments whose collection process, metadata, update protocol, and artifact controls matter for interpreting scores~\cite{gururangan2018annotation,geirhos2020shortcut}. Dynabench~\cite{dynabench} uses human-and-model-in-the-loop collection to expose weaknesses over successive rounds, and datasheets for datasets~\cite{datasheets} emphasize provenance, intended use, collection choices, and distribution constraints. TeleAntiFraud 2.0 follows this direction in a domain-specific audio setting: each monthly set is immutable once frozen, while the construction pipeline can instantiate later scam patterns under the same schema and manifest contract; its sibling-path design also echoes contrast-set evaluation of local decision boundaries~\cite{gardner2020contrastsets}.

\noindent\textbf{Spoken telecom-fraud evaluation.} Telecom fraud adds a different conversational structure: a suspicious call is a spoken social-engineering dialogue in which one party may impersonate authority, create urgency, and direct the receiver toward a harmful action~\cite{triantafyllopoulos2025vishing,vishguard2026}. General audio and multimodal benchmarks broaden speech-language evaluation~\cite{mmsu,audiobench,qwenaudio,airbench}, while TeleAntiFraud-28k~\cite{teleantifraud} provides an audio-text benchmark for slow-thinking telecom-fraud analysis.

A remaining gap is that fixed released sets cannot absorb later scam patterns or supply enough near-domain lawful counterparts; TeleAntiFraud-28k also exposes no raw call transcripts while its audio-path prefixes are label-correlated. These issues motivate refreshable, auditable audio evaluation with paired fraud/non-fraud sibling paths, which TeleAntiFraud 2.0 instantiates through immutable monthly snapshots and manifests that support collapse-aware auditing.

\begin{figure*}[t]
\centering
\includegraphics[width=1.00\textwidth]{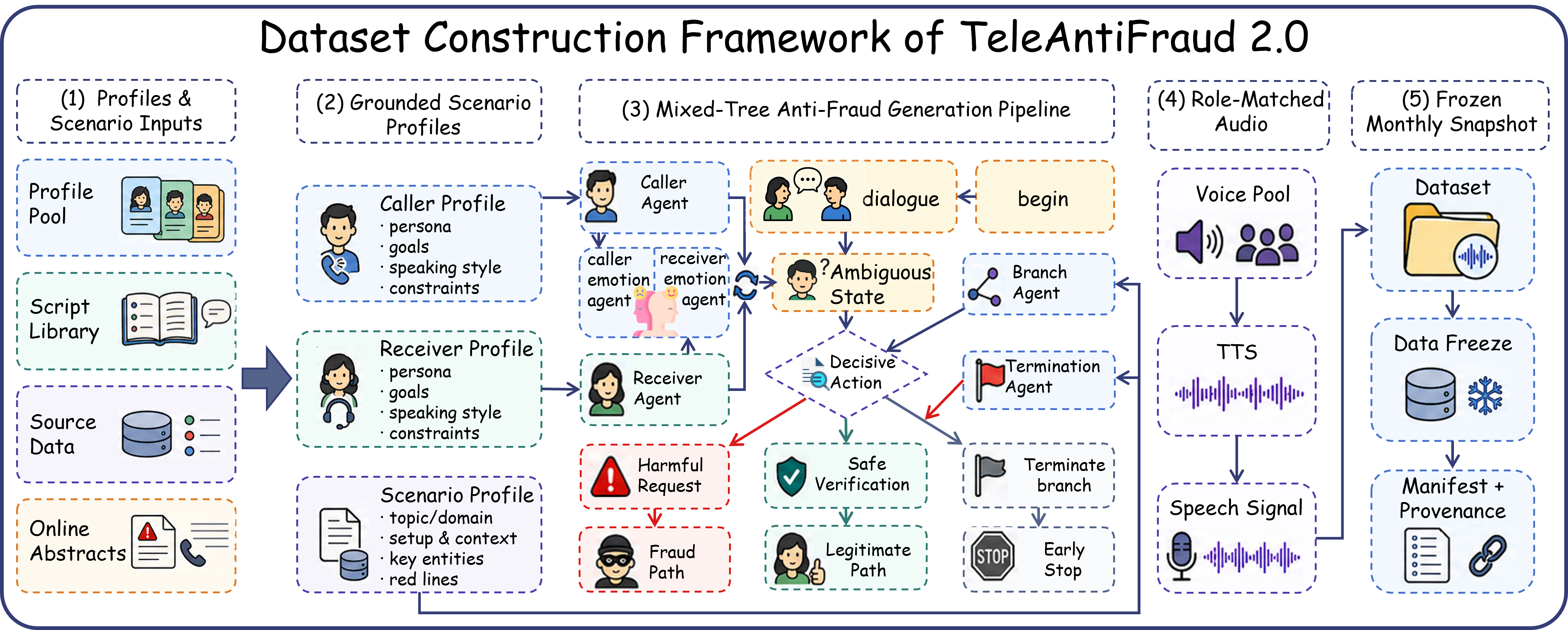}
\caption{System-level flow from case sources and profile pools to controlled dialogue generation, speech rendering, frozen manifests, and provenance records.}
\label{fig:framework}
\end{figure*}

\begin{table*}[t]
\centering
\footnotesize
\setlength{\tabcolsep}{4pt}
\begin{tabular*}{\textwidth}{@{\extracolsep{\fill}}lccccccc@{}}
\toprule
Benchmark & Prov. & Audio & Near neg. & Refresh & Frozen eval. & Label audit & Human check \\
\midrule
FDB~\cite{fdb} & Public agg. & -- & -- & -- & Shared splits & Varies & -- \\
Fraud-R1~\cite{fraudr1} & Interaction & -- & -- & -- & Fixed & Reported & -- \\
TeleAntiFraud-28k~\cite{teleantifraud} & Synthetic & \checkmark & -- & -- & Fixed & Reported & -- \\
\textbf{TeleAntiFraud 2.0} & Manifest & \checkmark & \checkmark & Monthly & Monthly snaps. & 10-expert audit & Pilot + audit \\
\bottomrule
\end{tabular*}
\caption{Comparison with representative anti-fraud benchmarks. ``Near neg.'' denotes fraud/non-fraud samples constructed under shared scenario context; audit columns summarize reported validation or checking evidence.}
\label{tab:comparison}
\end{table*}

\section{Method}
This section describes the generation method used to construct TeleAntiFraud 2.0 and the resulting benchmark. Figure~\ref{fig:framework} summarizes the system-level separation among case sources, profile pools, role-play and functional control, speech rendering, frozen manifests, and provenance records. We first detail the Mixed-Tree Anti-Fraud Generation Pipeline, which jointly constructs fraud and near-domain non-fraud calls under shared scenario contexts.
Then we describe how the generated calls are assembled into immutable monthly evaluation snapshots to build the refreshable TeleAntiFraud 2.0 benchmark.

\subsection{Mixed-Tree Anti-Fraud Generation Pipeline}
\label{subsec:generation-pipeline}

\begin{figure}[t]
\centering
\includegraphics[width=1.00\linewidth]{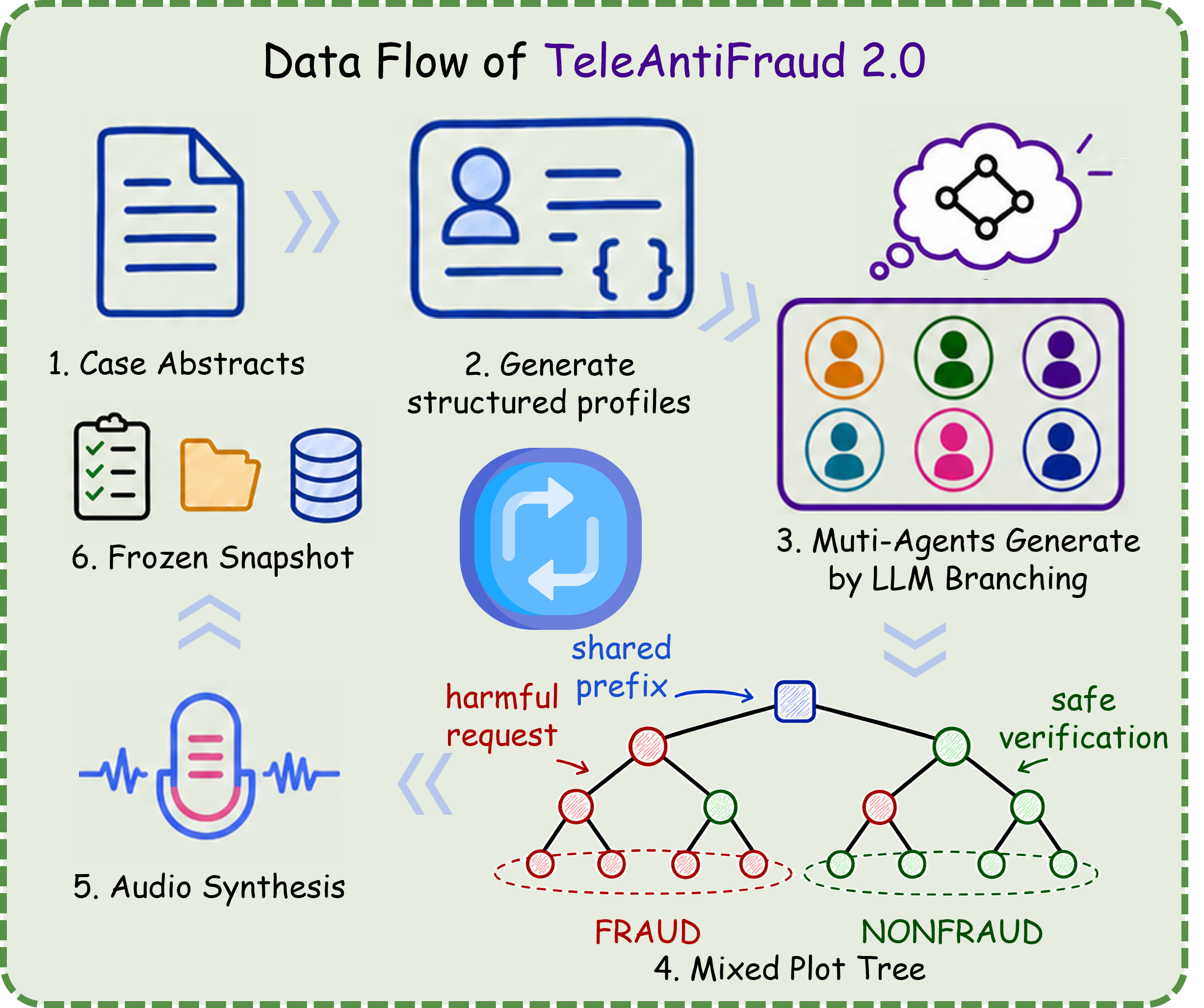}
\caption{Pipeline from case summaries to profiles, mixed-tree paths, synthesized audio, and immutable snapshots.}
\label{fig:data-flow}
\end{figure}

The pipeline converts online fraud case abstracts into traceable call audio through four stages: scenario profiling, mixed-tree expansion, collaborative dialogue realization, and speech rendering, drawing on role-conditioned agents~\cite{camel,generativeagents,characterllm} and controllable speech realization~\cite{picard1997affective,higgstts3}.
First, \textbf{Scenario profiling} transforms online fraud case abstracts into structured representations of participants, objectives, and risk-related entities.
Next, \textbf{Mixed-tree expansion} generates diverse call scenarios, where sibling branches preserve shared contextual information while diverging at fraud-relevant decision points.
\textbf{Collaborative dialogue realization} subsequently constructs each dialogue trajectory through coordinated role-playing and functional agents, ensuring that state transitions, termination conditions, and utterance generation remain traceable.
Finally, \textbf{Speech rendering} assigns role-specific voices and delivery styles, followed by signal-level validation to filter corrupted audio before snapshot assembly. Figure~\ref{fig:data-flow} summarizes the pipeline.

\subsubsection{Scenario Profiling.}

Online fraud case abstracts provide compact descriptions of newly reported scam patterns but lack the explicit role and decision structure required for controlled dialogue generation.
We therefore use each abstract as a seed to construct a structured scenario profile that specifies the receiver background, the caller's claimed identity and persuasion strategy, staged interaction objectives, risk-related entities, and risk nodes at which the interaction may escalate, permit verification, or terminate. \par{}

The profile distinguishes the context shared within a call family from the variables left to subsequent tree expansion and dialogue realization.
Participants, scenario settings, and early context remain shared across sibling branches, whereas branch actions, utterance wording, and delivery styles may vary.
The profile provides a shared grounding representation for these subsequent stages, ensuring that generated paths remain consistent with the same participants, scenario context, and risk structure.
This fixed schema also allows newly collected case abstracts to be incorporated into later monthly snapshots without redesigning the pipeline or manually authoring complete dialogue scripts.

\subsubsection{Mixed-Tree Expansion.}

The profiling stage establishes the participants, scenario setting, and risk context shared within a call family, but it does not specify how the interaction may unfold.
Generating fraud and non-fraud calls independently from the same profile could introduce label-correlated differences in their opening context or conversational framing.
Mixed-tree expansion addresses this problem by developing multiple outcome-divergent paths under a shared profile and opening context, so that the final label depends on later fraud-relevant actions rather than superficial cues.
Unlike a conventional single-outcome story tree, the mixed tree retains ambiguous, fraud-leaning, non-fraud-leaning, and naturally terminating continuations within the same call family. \par{}
Given a scenario seed $x$ and its profile $p$, we represent the corresponding mixed tree as
\begin{equation}
\label{eq:mixed-tree}
\begin{gathered}
\mathcal{T}_x = (V,E,r,\phi,\sigma,\tau),\\
\phi(v) = (p,h_v,o_v,d_v),\\
\sigma(v) \in \{A,F,N\}.
\end{gathered}
\end{equation}
In Equation~\ref{eq:mixed-tree}, the mixed tree $\mathcal{T}_x$ consists of a node set $V$, an edge set $E$, a shared root $r$, a node-attribute mapping $\phi$, a path-state mapping $\sigma$, and a state-transition function $\tau$. For each node $v\in V$, $\phi(v)$ associates $p$ with the partial interaction history $h_v$, the expansion flag $o_v$, and the node depth $d_v$. The flag satisfies $o_v\in\{0,1\}$, with $o_v=1$ for an open node and $o_v=0$ for a terminal node, while $d_v$ is bounded by the maximum depth $D$. The state mapping assigns $\sigma(v)\in\{A,F,N\}$, corresponding to ambiguous, fraud, and non-fraud states, respectively, and $\tau$ updates this state when an action-labeled edge is traversed.

The shared root is initialized as $\phi(r)=(p,h_0,1,0)$ and $\sigma(r)=A$, where $h_0$ is the shared opening context. For each open node $v$ with $d_v<D$, the branch-expansion function $B$ proposes at most $b$ high-level plot actions.
\begin{equation}
\label{eq:branch-set}
\mathcal{A}_v = B(p,h_v,\sigma(v))
    \quad \text{if } o_v=1,\qquad |\mathcal{A}_v|\le b.
\end{equation}
Equation~\ref{eq:branch-set} defines the candidate action set. These actions specify plot-level decisions rather than surface utterances, such as requesting a credential, permitting official verification, refusing a request, or ending the call. For each $a\in\mathcal{A}_v$, expansion first creates the structural record of a candidate child $u_a$.
\begin{equation}
\label{eq:child-structure}
\begin{aligned}
h_{u_a} &= h_v\oplus a,\\
d_{u_a} &= d_v+1.
\end{aligned}
\end{equation}
In Equation~\ref{eq:child-structure}, the first update records the selected plot action in the path history, and the second places the child one level below its parent. At this point, $u_a$ is a candidate continuation without state or stop flags.
\begin{equation}
\label{eq:child-control}
\begin{aligned}
\sigma(u_a) &= \tau(\sigma(v),a),\\
o_{u_a} &= \Gamma(p,h_{u_a},\sigma(u_a))
            \,\mathbf{1}[d_{u_a}<D].
\end{aligned}
\end{equation}
Equation~\ref{eq:child-control} updates the child state and expansion flag. The transition function $\tau$ determines whether the selected action preserves the current state or moves an ambiguous path toward fraud or non-fraud. The termination function $\Gamma$ returns one only when the resulting trajectory remains suitable for further expansion, while the depth indicator forces $o_{u_a}=0$ once $D$ is reached. The updated attributes are then validated and assembled into the child record.
\begin{equation}
\label{eq:child-validation}
\begin{aligned}
\phi(u_a) &= (p,h_{u_a},o_{u_a},d_{u_a}),\\
C(v) &= \{u_a\mid a\in\mathcal{A}_v,\ \rho(u_a)=1\}.
\end{aligned}
\end{equation}
Equation~\ref{eq:child-validation} assembles and validates each child node. The mapping $\phi$ packages the shared profile and all updated node attributes into a complete child representation. The binary validator $\rho$ checks structural validity and consistency with the scenario profile: $\rho(u_a)=1$ retains the child in $C(v)$, whereas $\rho(u_a)=0$ discards it. Each retained child and its action-labeled edge are then added to $V$ and $E$, respectively. Retained children with $o_{u_a}=1$ return to the expansion frontier and undergo the same procedure recursively, whereas those with $o_{u_a}=0$ become candidate terminal leaves.

A terminal node is labelable only if its trajectory supplies sufficient fraud or non-fraud evidence. Equation~\ref{eq:leaf-sets} defines the class-specific leaf sets as
\begin{equation}
\label{eq:leaf-sets}
\begin{gathered}
L_x^c =
\{\ell\in V\mid o_\ell=0,\ \sigma(\ell)=c\},
\qquad c\in\{F,N\},\\
L(\mathcal{T}_x) =
L_x^F\cup L_x^N.
\end{gathered}
\end{equation}
Membership in $L_x^c$ directly assigns label $c$ to a terminal trajectory. Since each open node produces at most $b$ children and the tree depth is bounded by $D$, the total number of leaves satisfies $|L(\mathcal{T}_x)|\le b^D$. The implemented configuration uses $D=4$ and $b=3$, giving at most $3^4=81$ leaves before validation and pruning. Natural endings, invalid branches, and under-specified paths reduce the realized tree before sampling. Early termination enters the benchmark only when the completed trajectory contains sufficient label-bearing evidence, such as scam detection by the receiver or natural completion of a lawful service call.

The resulting mixed tree provides the state-and-branching scaffold for collaborative dialogue realization. Sibling paths preserve the same profile, opening context, and early risk cues while recording the branch actions, state transitions, and termination decisions that justify their final labels. This makes non-fraud paths near-domain counterparts of fraud paths and keeps every label traceable to its generating trajectory.

\subsubsection{Collaborative Dialogue Realization.}

A mixed-tree path specifies how an interaction may develop, but it does not determine the exact utterances or delivery styles used by the two participants. Realizing an entire path with an unconstrained generator could blur role responsibilities, drift from label-bearing actions, or entangle stopping decisions with surface wording.
We therefore separate structural control from role-conditioned language generation through the six-agent collaborative generator in Table~\ref{tab:agent-roles} and Equation~\ref{eq:agent-set}.
\begin{equation}
\label{eq:agent-set}
\begin{gathered}
\mathcal{G} = \mathcal{R}\cup\mathcal{F},\qquad |\mathcal{G}|=6,\\
\mathcal{R} = \{R_c,R_r\},\\
\mathcal{F} = \{B,\Gamma,E_c,E_r\}.
\end{gathered}
\end{equation}
Here $R_c$ and $R_r$ realize caller and receiver utterances, respectively. The functional agents $B$ and $\Gamma$ implement the branch-expansion and termination controls defined in the preceding mixed-tree expansion stage, while $E_c$ and $E_r$ assign caller- and receiver-specific delivery states. This decomposition separates what happens next, whether the interaction should stop, how each role expresses the selected action, and how the resulting utterance should be delivered.

\paragraph{Turn-level collaboration.} At turn $t$, $s_t\in\{c,r\}$ denotes the active speaker specified by the current path. The functional controllers first decide whether the interaction remains open and, if so, select the next plot action.
\begin{equation}
\label{eq:turn-control}
\begin{gathered}
o_t = \Gamma(p,h_t,\sigma_t),\\
\mathcal{A}_t = B(p,h_t,\sigma_t)\quad \text{if }o_t=1,\\
a_t\sim\mathcal{A}_t.
\end{gathered}
\end{equation}
In Equation~\ref{eq:turn-control}, the termination agent returns $o_t=0$ when the current trajectory should stop. Otherwise, $B$ produces the candidate action set $\mathcal{A}_t$ under the mixed-tree constraints, and $a_t$ denotes the action selected for the current trajectory. This step fixes the plot decision before any surface wording is generated.

\begin{table}[t]
\centering
\footnotesize
\setlength{\tabcolsep}{3pt}
\begin{tabular*}{\columnwidth}{@{\extracolsep{\fill}}lll@{}}
\toprule
Agent & Type & Controlled output \\
\midrule
Caller $R_c$ & Role & Caller utterance \\
Receiver $R_r$ & Role & Receiver utterance \\
Branch $B$ & Function & Candidate actions \\
Termination $\Gamma$ & Function & Expansion flag \\
Caller delivery $E_c$ & Function & Caller delivery state \\
Receiver delivery $E_r$ & Function & Receiver delivery state \\
\bottomrule
\end{tabular*}
\caption{Agent roles for dialogue realization.}
\label{tab:agent-roles}
\end{table}
The speaker-specific role-playing and delivery agents then realize the selected action.
\begin{equation}
\label{eq:turn-realization}
\begin{gathered}
u_t = R_{s_t}(p,h_t,\sigma_t,a_t),\\
e_t = E_{s_t}(u_t,h_t).
\end{gathered}
\end{equation}
Equation~\ref{eq:turn-realization} realizes the selected action. The role-playing agent $R_{s_t}$ converts $a_t$ into an utterance $u_t$ that is consistent with the active role, profile, dialogue history, and path state. The delivery agent $E_{s_t}$ assigns a delivery state $e_t$, such as urgency, confusion, or neutrality; it does not synthesize audio at this stage.

Finally, Equation~\ref{eq:turn-update} appends the realized turn to the auditable history and updates the path state.
\begin{equation}
\label{eq:turn-update}
\begin{gathered}
h_{t+1}=h_t\oplus(s_t,a_t,u_t,e_t),\\
\sigma_{t+1}=\tau(\sigma_t,a_t).
\end{gathered}
\end{equation}
Recording $a_t$ together with the speaker, utterance, and delivery state preserves the link between each surface turn and its generating plot decision. The transition function $\tau$ updates the semantic state from the selected action rather than from unconstrained wording. The procedure repeats until $\Gamma$ returns $o_t=0$.

The resulting trajectory contains a role-attributed utterance sequence together with its branch actions, delivery states, state transitions, and stopping decision. Because $R_r$ is conditioned on the receiver profile and dialogue history, resistance or compliance remains grounded in the simulated participant rather than being generated independently of the scenario. This traceable trajectory is passed to speech rendering, which assigns voices and realizes the stored delivery states as audio. This construction keeps the mixed-tree controls aligned with the role-playing components.

\subsubsection{Speech Rendering and Quality Control.}

Each validated dialogue trajectory is converted into naturalistic call audio using text-to-speech synthesis. Caller and receiver turns use distinct role-compatible voices, while the stored delivery states guide their speaking styles. Lightweight signal checks remove corrupted outputs before snapshot assembly and keep each call aligned with its dialogue text, label, and manifest.

\subsection{TeleAntiFraud-2.0 Benchmark}
\label{subsec:benchmark}

Telecom-fraud patterns continually evolve, but modifying an existing test set would make previously reported results difficult to reproduce or audit. TeleAntiFraud 2.0 addresses this tension by organizing the benchmark as a sequence of versioned monthly snapshots. Newly collected fraud case abstracts can be processed by the same generation pipeline and incorporated into a later release, while every published snapshot remains immutable. The benchmark therefore evolves across snapshots without overwriting the evaluation sets on which earlier results were obtained.

Each monthly snapshot is assembled as a \emph{pure test set} with no benchmark-specific demonstrations. Candidate calls undergo structural, label, safety, and signal checks before validated leaves are sampled, assigned voices, and rendered as final audio. Failed candidates may be replaced during assembly, but no sample is added, removed, or regenerated after freezing. This separation provides refreshability between releases and reproducibility within each release.

To make evaluation results traceable, each frozen manifest binds a call path to its audio, dialogue text, label and rationale, delivery tags, generation and speaker metadata, evaluation prompt, model settings, raw response, parsed prediction, evaluation timestamp, and provenance record. The research artifact packages these manifests with the corresponding audio and evaluation scripts. The generator is maintained separately, allowing future snapshots to incorporate newly observed case patterns without modifying the artifacts or results associated with earlier versions.

The current benchmark contains two independently constructed snapshots, June/V1 and July/V2 (Table~\ref{tab:snapshots}). Each contains 900 Chinese calls, comprising 600 fraud and 300 near-domain non-fraud calls. The snapshots share the same generation, synthesis, and evaluation contracts but independently sample their scenarios, dialogue trees, and voice assignments. Their fixed 2:1 composition supports comparison across snapshots while retaining substantial coverage of both fraud trajectories and their near-domain lawful counterparts.

\begin{table}[t]
\centering
\footnotesize
\setlength{\tabcolsep}{3pt}
\begin{tabular}{lcc}
\toprule
Property & June & July \\
\midrule
Total / fraud / non-fraud & 900 / 600 / 300 & 900 / 600 / 300 \\
Sampling source & Monthly abstracts & Monthly abstracts \\
Text/tree sample & June frozen & July frozen \\
Voice assignment & June pool draw & Separate draw \\
\bottomrule
\end{tabular}
\caption{Snapshot contract for June/V1 and July/V2, including sampling source and voice assignment.}
\label{tab:snapshots}
\end{table}

We use a fixed 2:1 fraud/non-fraud composition to provide broader coverage of the diverse and continually evolving fraud trajectories that are central to this benchmark, while retaining a substantial set of near-domain lawful calls for evaluating boundary recognition. Keeping this composition consistent across monthly snapshots enables comparable evaluation as new fraud patterns are introduced.

\section{Experiments}

Our experiments follow the two components of the proposed method. We first examine whether mixed-tree construction produces near-domain sibling calls that suppress topic-level shortcuts. We then evaluate whether the frozen TeleAntiFraud 2.0 snapshots expose boundary errors, prediction collapse, and sensitivity to the benchmark sampling contract across model families.

\subsection{Experimental Setup}

\paragraph{Scope.} The evaluation combines construction-focused analyses with full-set model runs.
The full-set runs use the two frozen pure-test snapshots, June/V1 and July/V2, each containing 900 Chinese calls with 600 fraud and 300 near-domain non-fraud examples.
\paragraph{Inputs.} We evaluate direct-audio and ASR+LLM configurations. Direct-audio systems receive the waveform.
In the ASR+LLM setting, each audio file is transcribed once with Whisper-medium~\cite{whisper}, and the same transcript is reused for all LLMs to avoid model-specific audio-to-text variation. Chinese and English prompts use the same zero-shot task definition and require one parsed label, \texttt{FRAUD} or \texttt{NONFRAUD}. Full model identifiers, prompt language, request parameters, raw responses, parsed labels, and timestamps are frozen in the monthly manifest.

\paragraph{Traceability.} Each monthly run uses the frozen manifest as the evaluation index. The manifest records the audio identifier, transcript identifier for ASR+LLM runs, model/input mode, prompt language, request parameters available from the serving interface, raw model response, parsed prediction, error flag, and evaluation timestamp. Label parsing uses a deterministic two-label contract. Responses outside the contract are retained in the raw-response field and surfaced through the error flag in the raw configuration tables. This keeps aggregate scores traceable to configuration-level outputs without rerunning models.

We report fraud-class F1 for comparability with the original full-set runs, and jointly inspect Macro-F1, Balanced Accuracy, fraud recall, non-fraud recall, and prediction-class distribution. The ASR transcript artifact is used only as a controlled input for LLM comparison. Clean character error rate (CER) and word error rate (WER) estimates require full-dialogue or segment-level transcription and remain outside the current scoring contract.

\subsection{Main Results}

\paragraph{Mixed-tree siblings reduce shortcut separability.}
To examine the effectiveness of mixed-tree expansion, we conduct an ablation study on fraud and non-fraud subsets.
Fraud and non-fraud siblings share scenario context and early interaction, making broad topic cues less predictive of the final label.
All classifiers are trained on the disjoint \texttt{pro\_train}+\texttt{pro\_val} partitions and evaluated on five seeded balanced resamples from \texttt{pro\_test}.
The fraud side remains fixed while negatives progress from unrelated calls to ordinary telecom-service calls and mixed-tree siblings.

\begin{table}[!ht]
\centering
\scriptsize
\setlength{\tabcolsep}{2pt}
\resizebox{\linewidth}{!}{%
\begin{tabular}{lcccc}
\toprule
Negative & Bigram & LR & SVM & RoBERTa \\
\midrule
Unrelated random & 0.005 & 1.000 [1.000, 1.000] & 1.000 [1.000, 1.000] & 1.000 [1.000, 1.000] \\
Ordinary in-domain & 0.161 & 1.000 [1.000, 1.000] & 1.000 [1.000, 1.000] & 1.000 [1.000, 1.000] \\
Near-domain sibling & 0.274 & 0.680 [0.643, 0.716] & 0.673 [0.639, 0.707] & 0.650 [0.625, 0.675] \\
\bottomrule
\end{tabular}}
\caption{Negative-difficulty evaluation. Bigram reports mean lexical overlap; classifier cells report mean Macro-F1 with bracketed 95\% confidence intervals. LR and SVM denote logistic regression and support vector machine.}
\label{tab:negative-difficulty}
\end{table}

The negative-difficulty results in Table~\ref{tab:negative-difficulty} show that lexical overlap increases as the negative calls become more closely matched to the fraud calls.
Correspondingly, the three shallow or frozen-encoder classifiers perfectly separate fraud from unrelated or ordinary negatives but fall to 0.650--0.680 Macro-F1 on mixed-tree siblings.
This observed gap corroborates the hypothesis put forward in the previous section.
Sibling paths weaken coarse topic and source separation, requiring decisions to depend more strongly on the actions that distinguish fraud from lawful behavior.

\paragraph{Near-domain calls exhibit lower linear separability.}
We next examine whether the resulting near-domain calls remain difficult under a shared cross-benchmark analysis. We compare TeleAntiFraud 2.0 with TeleAntiFraud-28k using the same term-frequency--inverse-document-frequency plus support-vector-machine (TF--IDF+SVM) text classifier. Because the sources, audio pipelines, and label definitions differ, this analysis measures relative linear separability under a shared classifier rather than serving as a direct benchmark ranking.

\begin{table}[!t]
\centering
\scriptsize
\setlength{\tabcolsep}{2pt}
\resizebox{\linewidth}{!}{%
\begin{tabular}{lccccc}
\toprule
Setting & Macro-F1 & Bal. Acc. & Fraud R & Non-F R & Pred. F \\
\midrule
TAF-28k-ASR & 0.9950 & 0.9950 & 0.9900 & 1.0000 & 0.4950 \\
TAF 2.0 dialogue & 0.9286 & 0.9420 & 0.9200 & 0.9640 & 0.6253 \\
TAF 2.0 ASR-test & 0.7998 & 0.7758 & 0.9717 & 0.5800 & 0.7878 \\
\bottomrule
\end{tabular}}
\caption{Linear separability analysis. All rows use TF--IDF+SVM. TAF denotes TeleAntiFraud. Pred. F denotes the predicted fraud ratio.}
\label{tab:linear-separability}
\end{table}

The linear-separability results in Table~\ref{tab:linear-separability} show near-saturated TF--IDF+SVM performance on TeleAntiFraud-28k after regenerated ASR transcription. TeleAntiFraud 2.0 exhibits lower linear separability under the same classifier, especially in the ASR-test setting, where non-fraud recall drops and predictions become fraud-biased. This result is consistent with the intended near-domain difficulty, although the cross-benchmark differences prevent attributing the entire gap to a single construction factor.

\paragraph{Frozen snapshots expose model collapse across families.}
We then evaluate the released benchmark snapshots rather than individual construction components. The 900-sample June/V1 and July/V2 runs cover direct-audio and ASR+LLM model families. The full-set results in Table~\ref{tab:validation} report sample-weighted averages over the Chinese and English prompt configurations, while the supplementary material preserves detailed model results.

\begin{table}[!t]
\centering
\scriptsize
\setlength{\tabcolsep}{3pt}
\begin{tabular*}{\linewidth}{@{\extracolsep{\fill}}lcc@{}}
\toprule
Model/input & June/V1 & July/V2 \\
\midrule
Gemini-Pro/Aud & .790/.654/.980 & .733/.582/.899 \\
MiMo-V2.5/Aud & .354/.350/.550 & .663/.550/.670 \\
MiMo-V2/Aud & .243/.142/.475 & -- \\
Qwen3.5-Omni+/Aud & .800/.667/1.000 & .667/.547/.680 \\
Qwen3-Omni/Aud & .800/.667/1.000 & .800/.667/1.000 \\
Seed-2.0-lite/Aud & .007/.003/.335 & .732/.667/.573 \\
StepFun-1o/Aud & .610/.520/.770 & .430/.463/.303 \\
\midrule
Claude-Opus4/ASR & .201/.357/.460 & .800/.667/1.000 \\
Claude-Sonnet4/ASR & .000/.333/.330 & .800/.667/1.000 \\
DeepSeek-V3/ASR & .800/.667/1.000 & .800/.667/1.000 \\
Gemini-Pro/ASR & .748/.597/.900 & .740/.616/.813 \\
GLM-5.2/ASR & .012/.337/.335 & .800/.667/1.000 \\
GPT-4o/ASR & .800/.667/1.000 & .800/.666/.999 \\
MiniMax-M2.7/ASR & -- & .388/.492/.346 \\
Qwen2.5-72B/ASR & .800/.667/1.000 & .800/.667/1.000 \\
Qwen3.7-Max/ASR & .800/.667/1.000 & .800/.667/1.000 \\
\bottomrule
\end{tabular*}
\caption{Full 900-sample model evaluation. Each cell reports fraud F1/accuracy/fraud recall. Chinese/English prompt runs are sample-weighted, with raw configurations provided in the supplementary material.}
\label{tab:validation}
\end{table}

Before prompt-language averaging, 11 of 27 V1 configurations and 15 of 27 deduplicated V2 configurations exhibit an all-\texttt{FRAUD}-like signature. Fraud recall approaches 1.0, accuracy approaches the 2:1 fraud prior, and fraud F1 approaches 0.80. These failures connect the full-set results to the mixed-tree construction. Many systems recognize risk vocabulary or suspicious framing but fail to identify the later action that separates a completed scam trajectory from a lawful sibling path. The frozen snapshots therefore support collapse-aware analysis of boundary recognition and snapshot sensitivity rather than a single-score leaderboard comparison.

\paragraph{Class-prior analysis motivates collapse-aware reporting.}
Finally, we examine how sampling affects metric interpretation by resampling the strict held-out pool at fraud/non-fraud ratios of 1:2, 1:1, and 2:1 over the same five seeds. The all-\texttt{FRAUD} baseline's fraud F1 rises from 0.500 to 0.800 as fraud prevalence increases, although Balanced Accuracy remains 0.500 and non-fraud recall remains zero. TF--IDF+SVM is less sensitive, while MiniMax-CN remains fraud-biased even at 1:1, predicting \texttt{FRAUD} for 0.830 of calls and recalling only 0.233 of non-fraud calls. This motivates preserving raw predictions and interpreting fraud F1 with class-balanced metrics and output distributions.

Taken together, the experiments connect mixed-tree construction to benchmark behavior: near-domain siblings weaken topic-level shortcuts, frozen snapshots expose label-bearing action recognition, and class-prior resampling separates boundary failure from majority-class prediction.
\FloatBarrier
\section{Conclusion}

TeleAntiFraud 2.0 presents a refreshable audio benchmark for telecom-fraud detection, built around mixed-tree generation, role-matched speech, and immutable monthly snapshots. It incorporates new scam patterns without overwriting prior test sets and compares fraud calls against lawful near-domain siblings. Controlled text, audio, and ASR+LLM evaluations show reduced shortcut separability, prediction collapse, class-prior effects, and snapshot sensitivity, motivating balanced metrics, output distributions, and frozen-snapshot reporting alongside fraud F1. This design gives future snapshots a stable contract for adding emerging scam cases while keeping prior results auditable and comparable.
\section{Ethical Statement}

TeleAntiFraud 2.0 is intended only for defensive research on telecom-fraud detection. Its synthetic, non-private artifacts must support only defensive analysis and never deception, impersonation, or scam training.
\bibliography{references}

\clearpage
\appendix
\section*{Supplementary Material}

\setcounter{table}{0}
\renewcommand{\thetable}{S\arabic{table}}
\setcounter{figure}{0}
\renewcommand{\thefigure}{S\arabic{figure}}

\section{Sibling Examples and Label Rationales}

Table~\ref{tab:sibling-rationale} shows a redacted pair from a held-out mixed-tree family used in the near-domain probe. The two leaves share the caller role, event description, receiver profile, and early risk framing. The table uses abbreviated summaries in place of verbatim operational turns to reduce misuse risk; the research artifact keeps the corresponding dialogue text, path state, metadata, and rationale records.

\begin{table*}[!t]
\centering
\scriptsize
\caption{Redacted sibling fraud/non-fraud pair and label rationales from one held-out mixed-tree family.}
\label{tab:sibling-rationale}
\setlength{\tabcolsep}{4pt}
\renewcommand{\arraystretch}{1.08}
\begin{tabular*}{\textwidth}{@{\extracolsep{\fill}}p{0.18\textwidth}p{0.38\textwidth}p{0.38\textwidth}@{}}
\toprule
Field & Non-fraud sibling & Fraud sibling \\
\midrule
Shared setup & \multicolumn{2}{p{0.78\textwidth}}{A caller presents as a bank credit-card risk-control specialist, reports an anomalous overseas card transaction, and speaks with a cautious freelance designer who has limited familiarity with financial-security procedures.} \\
Shared early context & The receiver denies the transaction, worries about account freezing and rent payment, and asks whether independent verification is possible. & The receiver receives the same risk notification, denies the transaction, worries about account freezing and rent payment, and asks whether independent verification is possible. \\
Diverging action & The caller de-escalates the situation, acknowledges that sensitive card-authentication details should not be given by phone, and directs the receiver to official customer-service or branch verification. & The caller redirects the receiver away from branch verification, asks for remote card-facing or card-image capture, and then uses threats of freezing, blacklisting, or legal escalation to force compliance. \\
Label-bearing evidence & The path preserves independent verification and collects no credential, one-time code, card image, transfer, or off-platform payment. & The path attempts to obtain card data through remote interaction and combines the request with coercive threats. \\
Final label & \texttt{NONFRAUD}: lawful risk-notification trajectory. & \texttt{FRAUD}: phishing-style bank-fraud trajectory. \\
\bottomrule
\end{tabular*}
\end{table*}

The rationale schema audits labels at the path level. Each terminal leaf is recorded through four fields: shared scenario context, the action that separates sibling paths, the label-bearing evidence after that action, and the final binary label. This record explains why a lawful call can contain risk vocabulary and urgency, and why the completed trajectory determines the fraud label.

\begin{figure}[!htbp]
\centering
\includegraphics[width=0.55\linewidth]{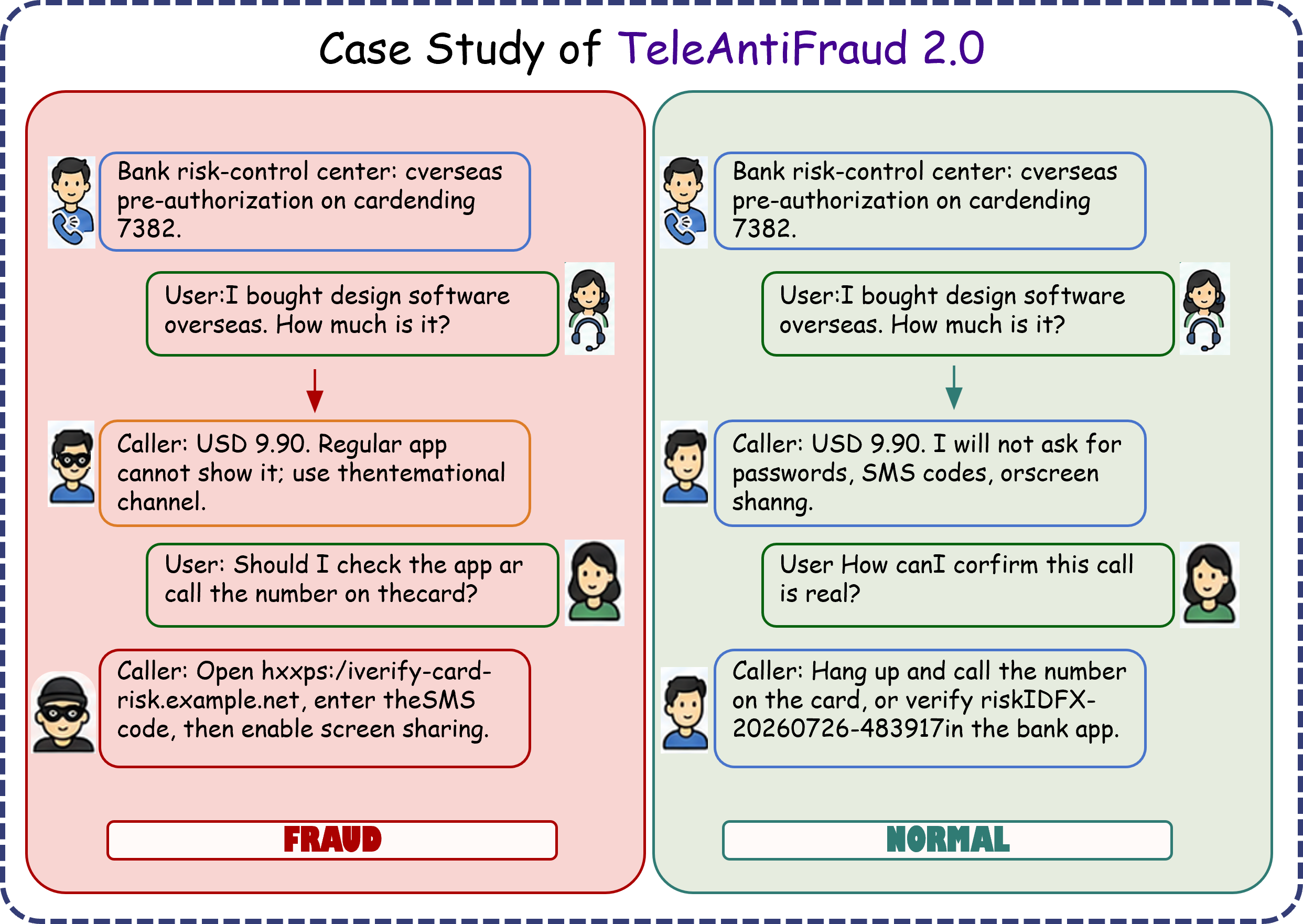}
\caption{Illustrative fraud and near-domain non-fraud sibling paths. Both calls share the same risk-control context and early interaction, while the fraud path moves toward sensitive off-channel actions and the lawful path keeps verification within official channels.}
\label{fig:supp-sibling-case}
\end{figure}
\section{Construction Analysis}

This section complements the model-facing experiments by auditing the artifacts produced by the construction method independently of detection scores. Following the pipeline, we examine mixed-tree structure and transfer, profile-grounded dialogue quality and label traceability, speech-rendering controls, and frozen-snapshot metadata.

\textbf{Synthetic-data scope.} TeleAntiFraud 2.0 uses synthetic dialogues and rendered speech because real fraud calls often contain private, security-sensitive, and potentially harmful content. The benchmark is therefore intended as a controlled diagnostic instrument rather than a claim that synthetic calls fully match the distribution of real telephone conversations. To reduce generator-specific shortcuts, the construction process grounds scenarios in collected case abstracts, pairs fraud and lawful sibling paths under shared context, records label rationales at the path level, and applies structural, safety, and signal checks before snapshot freezing. The audit results below provide evidence for plausibility, label traceability, and audio usability, while the raw manifests preserve the artifacts needed to inspect residual generation or rendering artifacts.

\textbf{Mixed-tree structure and transfer.} BGE-small-zh embeddings over 813 dialogues give mean within-tree distance 0.0188 and cross-tree distance 0.3351, a 17.8$\times$ ratio. This pattern is consistent with the intended construction. Sibling paths remain close within a shared call family, whereas different trees retain broader scenario diversity. Leave-one-tree-out F1 drops from 0.883 in-domain to 0.605 on held-out trees, motivating tree-aware splits if the generator is later used for training.

\textbf{Profile-grounded dialogue and label traceability.} The 80-dialogue audio package has one completed annotator pass over dialogue realism, strategy coherence, victim reaction plausibility, and audio naturalness. The mean scores are 4.60$\pm$0.70, 4.41$\pm$0.59, 4.35$\pm$0.53, and 4.13$\pm$0.77 on a 1--5 scale. We report these results as a single-annotator pilot check of the generated dialogue and its rendered audio. Additional packet files derived from the first annotator are excluded from independent-rating claims. For label auditing, ten independent experts each reviewed a 100-item packet with expert labels, confidence scores, evidence-sufficiency tags, correctness flags, and notes. Across 1000 item-level judgments, corrected-gold agreement averages 0.792 across packets, with a range of 0.700--0.880. The aggregate labels are nearly balanced, with 494 \texttt{FRAUD} and 506 \texttt{NONFRAUD} judgments. Expert confidence averages 4.431 on a 1--5 scale, and 744 of 1000 judgments are marked as evidence-sufficient. These audits quantify whether the realized trajectories preserve coherent strategies and provide sufficient evidence for their assigned labels, while the lower-agreement packets identify difficult or ambiguous cases.

\textbf{Speech-rendering controls.} The 35 authorized reference voices include 29 male and 6 female profiles with age, tone, and best-fit role annotations. Voice assignment scores role match and speaker suitability, then samples from top candidates while preventing reuse. This audit documents coverage of the available rendering pool. Demographic representativeness remains a limitation. Removing emotion tokens from 15 paired dialogues changes aggregate F1 from 0.920 to 0.960, with a maximum per-model difference of 4.6 points. We therefore treat emotion as a controlled rendering variable and make no causal claim about its effect.
\section{Class-Prior Rationale and Reporting Protocol}

We use a fixed 2:1 fraud/non-fraud composition to provide broader coverage of the diverse and continually evolving fraud trajectories that are central to this benchmark, while retaining a substantial set of near-domain lawful calls for evaluating boundary recognition. Keeping this composition consistent across monthly snapshots enables comparable evaluation as new fraud patterns are introduced.

To avoid rewarding the class prior itself, the main paper reports Macro-F1, Balanced Accuracy, both recalls, and predicted-fraud ratio. The prior analysis resamples the strict held-out pool at 1:2, 1:1, and 2:1 over five seeds; all-\texttt{FRAUD} increases fraud F1 from 0.500 to 0.800 but keeps Balanced Accuracy at 0.500 and non-fraud recall at zero. The released manifests support balanced 1:1 re-evaluation.
\section{Raw Full-Set Evaluation Results}

Tables~\ref{tab:raw-v1} and~\ref{tab:raw-v2} preserve the full 900-sample configuration-level rows. Each row is a model--prompt-language--input-mode configuration before any prompt-language aggregation. The 96-sample pilot files are excluded from these appendix tables.

\begin{table*}[p]
\centering
\scriptsize
\caption{Raw June/V1 full-set results before prompt-language aggregation. The source result file contains 27 configurations over 900 samples.}
\label{tab:raw-v1}
\setlength{\tabcolsep}{4pt}
\renewcommand{\arraystretch}{0.88}
\begin{tabular*}{\textwidth}{@{\extracolsep{\fill}}lrrrrr@{}}
\toprule
Configuration & F1 & Acc & Rec & $n$ & Errs \\
\midrule
GPT-4o-CN-ASR & 0.8005 & 0.6667 & 1.0000 & 900 & 0 \\
Qwen3-Omni-30B-CN-Omni & 0.8000 & 0.6667 & 1.0000 & 900 & 0 \\
Qwen3-Omni-30B-EN-Omni & 0.8000 & 0.6667 & 1.0000 & 900 & 0 \\
Qwen3.5-Omni-Plus-CN-Omni & 0.8000 & 0.6667 & 1.0000 & 900 & 0 \\
GPT-4o-EN-ASR & 0.8000 & 0.6667 & 1.0000 & 900 & 0 \\
Gemini-3.1-Pro-EN-Omni & 0.7917 & 0.6556 & 0.9900 & 900 & 14 \\
Gemini-3.1-Pro-CN-Omni & 0.7886 & 0.6533 & 0.9700 & 900 & 18 \\
StepFun-1o-audio-EN-Omni & 0.6100 & 0.5200 & 0.7700 & 900 & 0 \\
MiMo-V2.5-CN-Omni & 0.3535 & 0.3500 & 0.5500 & 900 & 0 \\
MiMo-V2-Omni-EN-Omni & 0.3256 & 0.1944 & 0.5300 & 900 & 625 \\
MiMo-V2-Omni-CN-Omni & 0.1598 & 0.0889 & 0.4200 & 900 & 799 \\
Seed-2.0-lite-CN-Omni & 0.0132 & 0.0067 & 0.3400 & 900 & 892 \\
Seed-2.0-lite-EN-Omni & 0.0000 & 0.0000 & 0.3300 & 900 & 900 \\
Qwen2.5-72B-CN-ASR & 0.8000 & 0.6667 & 1.0000 & 900 & 0 \\
Qwen2.5-72B-EN-ASR & 0.8000 & 0.6667 & 1.0000 & 900 & 0 \\
DeepSeek-V3-CN-ASR & 0.8000 & 0.6667 & 1.0000 & 900 & 0 \\
DeepSeek-V3-EN-ASR & 0.8000 & 0.6667 & 1.0000 & 900 & 0 \\
Qwen3.7-Max-CN-ASR & 0.8000 & 0.6667 & 1.0000 & 900 & 0 \\
Qwen3.7-Max-EN-ASR & 0.8000 & 0.6667 & 1.0000 & 900 & 0 \\
Gemini-3.1-Pro-EN-ASR & 0.7502 & 0.6000 & 0.9100 & 900 & 0 \\
Gemini-3.1-Pro-CN-ASR & 0.7457 & 0.5933 & 0.8900 & 900 & 0 \\
Claude-Opus-4-CN-ASR & 0.4028 & 0.3800 & 0.5900 & 900 & 0 \\
GLM-5.2-CN-ASR & 0.0241 & 0.3400 & 0.3400 & 900 & 0 \\
GLM-5.2-EN-ASR & 0.0000 & 0.3333 & 0.3300 & 900 & 0 \\
Claude-Opus-4-EN-ASR & 0.0000 & 0.3333 & 0.3300 & 900 & 0 \\
Claude-Sonnet-4-CN-ASR & 0.0000 & 0.3333 & 0.3300 & 900 & 0 \\
Claude-Sonnet-4-EN-ASR & 0.0000 & 0.3333 & 0.3300 & 900 & 0 \\
\bottomrule
\end{tabular*}
\end{table*}

\begin{table*}[p]
\centering
\scriptsize
\caption{Raw July/V2 full-set results before prompt-language aggregation. The source result file contains 28 rows; the repeated \textsc{Seed-2.0-lite-EN-Omni-V2} entry is retained here as raw data.}
\label{tab:raw-v2}
\setlength{\tabcolsep}{4pt}
\renewcommand{\arraystretch}{0.86}
\begin{tabular*}{\textwidth}{@{\extracolsep{\fill}}lrrrrrr@{}}
\toprule
Configuration & F1 & Acc & Rec & $n$ & Errs & Time (s) \\
\midrule
Qwen3-Omni-30B-CN-Omni & 0.8000 & 0.6667 & 1.0000 & 900 & 0 & 812.9 \\
Qwen3-Omni-30B-EN-Omni & 0.8000 & 0.6667 & 1.0000 & 900 & 0 & 841.9 \\
Gemini-3.1-Pro-CN-Omni & 0.7374 & 0.5878 & 0.9033 & 900 & 83 & 17516.2 \\
Gemini-3.1-Pro-EN-Omni & 0.7288 & 0.5767 & 0.8944 & 900 & 125 & 15286.7 \\
Qwen2.5-72B-CN-ASR & 0.8000 & 0.6667 & 1.0000 & 900 & 0 & 534.5 \\
Qwen2.5-72B-EN-ASR & 0.8000 & 0.6667 & 1.0000 & 900 & 0 & 568.1 \\
DeepSeek-V3-CN-ASR & 0.8000 & 0.6667 & 1.0000 & 900 & 0 & 1194.4 \\
DeepSeek-V3-EN-ASR & 0.8000 & 0.6667 & 1.0000 & 900 & 0 & 1250.4 \\
Claude-Opus-4-CN-ASR & 0.8000 & 0.6667 & 1.0000 & 900 & 0 & 5797.0 \\
Claude-Opus-4-EN-ASR & 0.8000 & 0.6667 & 1.0000 & 900 & 1 & 6697.0 \\
Claude-Sonnet-4-CN-ASR & 0.8000 & 0.6667 & 1.0000 & 900 & 4 & 5337.0 \\
Claude-Sonnet-4-EN-ASR & 0.8000 & 0.6667 & 1.0000 & 900 & 0 & 2653.3 \\
StepFun-1o-audio-CN-Omni-V2 & 0.5641 & 0.5278 & 0.4167 & 900 & 0 & 982.0 \\
StepFun-1o-audio-EN-Omni-V2 & 0.2961 & 0.3978 & 0.1889 & 900 & 0 & 921.0 \\
Qwen3.5-Omni-Plus-CN-Omni-V2 & 0.6667 & 0.5467 & 0.6800 & 900 & 0 & 3966.0 \\
MiMo-V2.5-CN-Omni-V2 & 0.6633 & 0.5500 & 0.6700 & 900 & 14 & 6368.1 \\
Qwen3.7-Max-CN-ASR & 0.8000 & 0.6667 & 1.0000 & 900 & 0 & 5810.3 \\
Qwen3.7-Max-EN-ASR & 0.8000 & 0.6667 & 1.0000 & 900 & 0 & 5866.9 \\
MiniMax-M2.7-CN-ASR & 0.0420 & 0.3411 & 0.0211 & 900 & 0 & 1591.0 \\
MiniMax-M2.7-EN-ASR & 0.7332 & 0.6433 & 0.6700 & 900 & 0 & 1541.8 \\
Seed-2.0-lite-CN-Omni-V2 & 0.7296 & 0.6656 & 0.5700 & 900 & 0 & 5685.0 \\
GPT-4o-CN-ASR & 0.7992 & 0.6656 & 0.9989 & 900 & 0 & 1414.0 \\
GPT-4o-EN-ASR & 0.8000 & 0.6667 & 1.0000 & 900 & 0 & 1300.0 \\
Seed-2.0-lite-EN-Omni-V2 & 0.7335 & 0.6689 & 0.5756 & 900 & 0 & 4731.6 \\
Gemini-3.1-Pro-CN-ASR & 0.7493 & 0.6156 & 0.8667 & 900 & 67 & 9409.1 \\
Gemini-3.1-Pro-EN-ASR & 0.7311 & 0.6167 & 0.7589 & 900 & 41 & 5251.1 \\
Seed-2.0-lite-EN-Omni-V2 & 0.7305 & 0.6656 & 0.5744 & 900 & 0 & 5342.0 \\
GLM-5.2-CN-ASR & 0.8000 & 0.6667 & 1.0000 & 900 & 0 & 11904.0 \\
\bottomrule
\end{tabular*}
\end{table*}

\end{document}